\documentclass[aps,twocolumn,groupedaddress,showpacs,amsmath,amssymb]{revtex4}

\usepackage{graphicx}
\usepackage{dcolumn}
\usepackage{bm}

\begin{document}

\title{Superconducting phase diagram and nontrivial band topology of structurally modulated Sn$_{1-x}$Sb$_{x}$}

\date{\today}
\author{Bin Liu$^{1,2}$}
\author{Chengcheng Xiao$^{3}$}
\author{Qinqing Zhu$^{4}$}
\author{Jifeng Wu$^{1,2}$}
\author{Yanwei Cui$^{1,5}$}
\author{Hangdong Wang$^{4}$}
\author{Zhicheng Wang$^{5}$}
\author{Yunhao Lu$^{5}$}
\email{luyh@zju.edu.cn}
\author{Zhi Ren$^{1}$}
\email{zhi.ren@wias.org.cn}
\author{Guang-han Cao$^{5}$}

\affiliation{$^{1}$School of Science, Westlake Institute for Advanced Study, Westlake University, Hangzhou 310064, P. R. China}
\affiliation{$^{2}$Department of Physics, Fudan University, Shanghai 200433, P. R. China}
\affiliation{$^{3}$State Key Laboratory of Silicon Materials, School of Materials Science and Engineering, Zhejiang University, Hangzhou 310027, P. R. China}
\affiliation{$^{4}$Department of Physics, Hangzhou Normal University, Hangzhou 310036, P. R. China}
\affiliation{$^{5}$Department of Physics, Zhejiang University, Hangzhou 310027, P. R. China}

\begin{abstract}
We report the discovery of superconductivity in binary alloy Sn$_{1-x}$Sb$_{x}$ with $x$ in the range of 0.43 to 0.6, which possesses a modulated rhombohedral structure due to the incommensurate ordering of Sn and Sb layers along the $c$-axis. The specific heat measurements indicate a weakly coupled, fully gapped superconducting state in this homogeneity range with a maximum bulk $T_{\rm c}$ of 1.58 K at $x$ = 0.46, though the electronic specific heat and Hall coefficients remain nearly $x$-independent. The nonmonotonic dependence of the bulk $T_{\rm c}$ is discussed in relation to the effects of Sb-layer intercalation between the [Sn$_{4}$Sb$_{3}$] seven-layer lamellae that are the essential building block for superconductivity. On the other hand, a zero-resistivity transition is found to take place well above the bulk superconducting transition, and the corresponding $T_{\rm c}$ increases monotonically with $x$ from 2.06 K to 3.29 K. This contrast, together with the uniform elements distribution revealed by energy dispersive x-ray mapping, implies that the resistive transition is due to the strain effect at the grain boundary rather than the compositional inhomogeneity.
The first-principles calculations on the representative composition Sn$_{4}$Sb$_{3}$ ($x$ = 0.43) indicate that it is topologically nontrivial similar to Sb,
but with different Z$_{2}$ invariants (0;111). Our results not only identify a new superconducting region in the Sn-Sb phase diagram, but also provide a viable platform to study the interplay between structural modulation, nontrivial band topology and superconductivity.
\end{abstract}

\maketitle
\maketitle

\section{Introduction}
Recently, alloys consisting of heavy elements have received much attention because of their strong spin-orbit coupling (SOC), which could give rise to a set of topological phases \cite{kaneRMP,QiRMP}, including topological semimetals (TSM) and topological (crystalline) insulators (TI).
One prominent example is the binary alloy made up of the semimetal Bi and the TSM Sb \cite{kaneRMP,QiRMP}. In particular, when $x$ falls in the
range between 0.07 and 0.22, a band gap opens due to band inversion in Bi$_{1-x}$Sb$_{x}$, which represents the first material realization of the three-dimensional TI \cite{paritycounting,Bi1-xSbxarpes,Bi1-xSbxalloy}.
Topological superconductors (TSC) are the superconducting analogues of TI, and have potential applications in fault-tolerant quantum computing \cite{QiRMP,AliRPP,AndoRPP}.
So far, however, the promising candidates for TSC are mostly compounds, either doped or stoichiometric, such as $A_{x}$Bi$_{2}$Se$_{3}$ ($A$ = Cu, Sr, Nb) \cite{CuxBi2Se3Hor,CuxBi2Se3sasaki,SrxBi2Se3,NbxBi2Se3}, Sn$_{1-x}$In$_{x}$Te \cite{Sn1-xInxTe}, Tl$_{5}$Te$_{3}$ \cite{Tl5Te3}, PbTaSe$_{2}$ \cite{PbTaSe2}, Au$_{2}$Pb\cite{Au2Pb}, Tl$_{0.6}$Bi$_{2}$Te$_{3}$\cite{Tl0.6Bi2Te3}, and $\beta$-Bi$_{2}$Pd\cite{PdBi2}.
It is therefore desirable to search for superconducting alloys similar to Bi$_{1-x}$Sb$_{x}$ as they might also be potential candidates for TSC with wide compositional tunability.

\begin{figure*}
\includegraphics*[width=12cm]{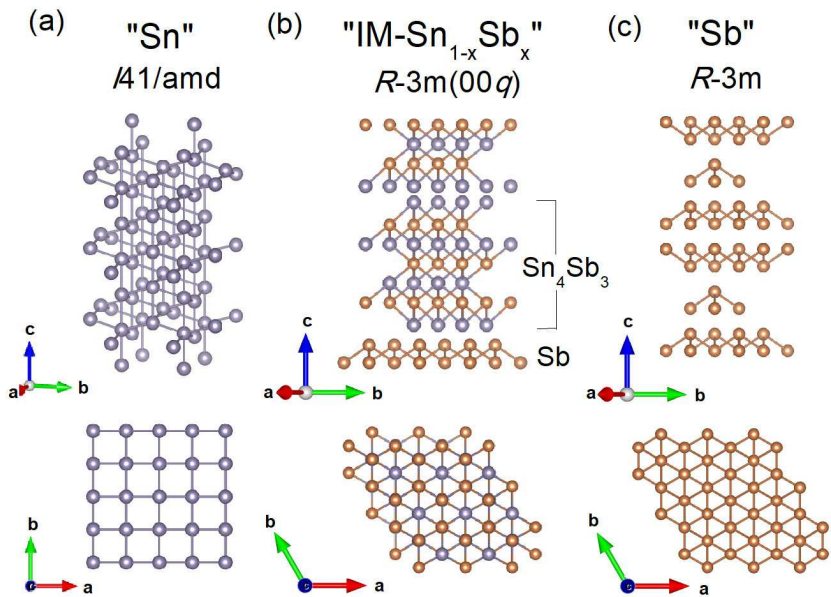}
\caption{(Color online)
Schematic structure of (a) Sn, (b) IM-Sn$_{1-x}$Sb$_{x}$, and (c) Sb parallel and perpendicular to the $c$-axis.
The building blocks in IM-Sn$_{1-x}$Sb$_{x}$ are labeled.
}
\label{fig1}
\end{figure*}
The binary Sn-Sb system appears to be attractive for this study. Its phase diagram has been studied extensively over the past few decades \cite{SnSbPD1}, and three single-phase regions are identified as a function of the Sb content $x$ \cite{SnSbPD2}: a solid solution of Sb in Sn for $x$ $\leq$ 0.08 [see Fig. 1(a)], incommensurately modulated (IM) Sn$_{1-x}$Sb$_{x}$ for 0.43 $\leq$ $x$ $\leq$ 0.6 [see Fig. 1(b)], and a solid solution of Sn in Sb for $x$ $\geq$ 0.92 [see Fig. 1(c)]. Unlike Bi$_{1-x}$Sb$_{x}$, where the Bi/Sb atoms are distributed disorderly in the lattice, IM-Sn$_{1-x}$Sb$_{x}$ is built up by seven-layer lamellae [Sn$-$Sb$-$Sn$-$Sb$-$Sn$-$Sb$-$Sn] and Sb layers stacked along the $c$-axis [Fig. 1(b)] \cite{SnSbPD2,SnSbstructure1,SnSbstructure2}. This leads to an incommensurate modulation of the rhombohedral structure, which can be described by the superspace group $R$$\overline{3}$$m$(00$q$) ($q$ is the modulation vector).
Intriguingly, the structure of pure Sb can also be characterized by the same superspace group, albeit with a commensurate $q$ value of 1.5 \cite{SnSbstructure2}.
Previous studies on the Sn-Sb system focus on its electrochemical performance as an anode material for Li-ion batteries \cite{SnSbEP1,SnSbEP2,SnSbEP3,SnSbEP4}, while little attention has paid to its physical properties.
It is until very recently that we have provided clear evidence that SnSb ($x$ = 0.5) is a bulk type-II superconductor below 1.50 K, while showing a zero resistivity transition at a significantly higher temperature of 2.48 K\cite{SnSbSC}.
For other $x$ values, however, no such study has been reported to date, and hence how the superconducting transitions evolve with $x$ remains unclear.
In addition, the topology of the electronic structure in these alloys has not been characterized.

In this paper, we report on the observation of SC over the whole homogeneity range of IM-Sn$_{1-x}$Sb$_{x}$ (0.43 $\leq$ $x$ $\leq$ 0.6) by means of resistivity, magnetic susceptibility and thermodynamic measurements on high quality polycrystalline samples with good homogeneity.
The resultant superconducting phase diagram shows that the bulk and resistive $T_{\rm c}$ evolve quite differently. The former exhibits a maximum of 1.58 K at $x$ = 0.46, though the electronic specific heat and low temperature Hall coefficients are nearly $x$ independent.
Furthermore, the analysis of the specific heat jump points to a weak coupling SC with nearly the same full gap structure for these samples. By contrast, a monotonic increase with $x$ is observed for the resistive $T_{\rm c}$, which is likely due to the strain effect developed at the grain boundaries. In addition, first-principles calculations show that the most Sn-rich composition $x$ = 0.43, whose structure consists solely of the [Sb$_{4}$Sn$_{3}$] septuple layers, belongs to the (0;111) Z$_{2}$ class and can be viewed as a weak TI with bent bands, suggesting that it may host Majorana fermions along the dislocation lines. The implication of this result on the band topology of other compositions is also discussed.

\section{Experimental}
Polycrystalline samples of IM-Sn$_{1-x}$Sb$_{x}$ were synthesized by the two-step method as described previously \cite{SnSbSC}. High-purity shots of Sn (99.99\%) and Sb (99.999\%) were mixed according to the stoichiometries of $x$ = 0.43, 0.46, 0.49, 0.53, 0.57, 0.6 in an argon-filled glove box, sealed in evacuated quartz tubes, and melted at 900 $^{\circ}$C for 2 days with intermittent shaking to ensure homogeneity, followed by direct quenching into cold water. The solidified ingots were then annealed without breaking the vacuum at 280-320 $^{\circ}$C for another 2 days and quenched again into cold water.
The crystal structure of the resulting samples was checked by powder x-ray diffraction (XRD) by using a PANalytical x-ray diffractometer with a monochromatic Cu-K$_{\alpha1}$ radiation at room temperature.
The data were collected with a step-scan mode in the 2$\theta$ range from 10$^{\circ}$ to 100$^{\circ}$. The Rietveld refinements of the XRD patterns were performed using the programm JANA2006 \cite{JANA}.
The morphology of the samples was investigated by Zeiss Supratm 55 schottky field emission scanning electron microscope (SEM), and their chemical compositions and homogeneity were measured with an energy-dispersive x-ray (EDX) spectrometer (Model Octane Plus).

For the sake of consistency, all the physical property measurements were carried out on samples obtained from the same ingot for each $x$. The resistivity was measured on bar-shaped samples by the standard four-probe method. The Hall coefficient was measured by sweeping the magnetic field from -9 T to 9 T. Resistivity and Hall coefficient measurements were performed in a Quantum Design physical property measurement system (PPMS-9 Dynacool)
. Specific heat measurements were done in a PPMS-9 Dynacool down to 1.8 K and in a PPMS-9 Evercool II down to 0.5 K. The two sets of data are consistent within 5\% in the overlapped temperature range. The measurements of dc magnetization down to 0.4 K were done on crushed powders in a Quantum Design magnetic property measurement system (MPMS3).

First principles calculations were carried out under the Density Functional Formalism as implemented in Vienna Ab-initio Simulation Package (VASP) \cite{VASP}. For exchange-correlation approximation, we used a generalized gradient approximation (GGA) with a projector augmented wave pseudopotentials \cite{pseudopotential}. The 5$s$5$p$ electrons for Sn and Sb were treated as valence electrons, and a 7$\times$7$\times$7 Monkhorst-Pack $k$-point mesh was employed to sample the Brilliouin zone of the rhombohedral unit cell \cite{mesh}. The energy convergence criterion was set to 10$^{-8}$ eV and the ionic geometry was optimized until the Hellmann-Feynman force acting on ions was less than 10$^{-3}$ eV/{\AA}.
Full relativistic effects and SOC were considered within the pseudopotential scheme.
The Z$_{2}$ topological invariants have been calculated using two different methods: the parity counting method \cite{paritycounting} and Wilson loop method\cite{wilsonloop}.

\section{Results and Discussion}
\subsection{X-ray diffraction and SEM characterization}
\begin{figure}
\includegraphics*[width=8.6cm]{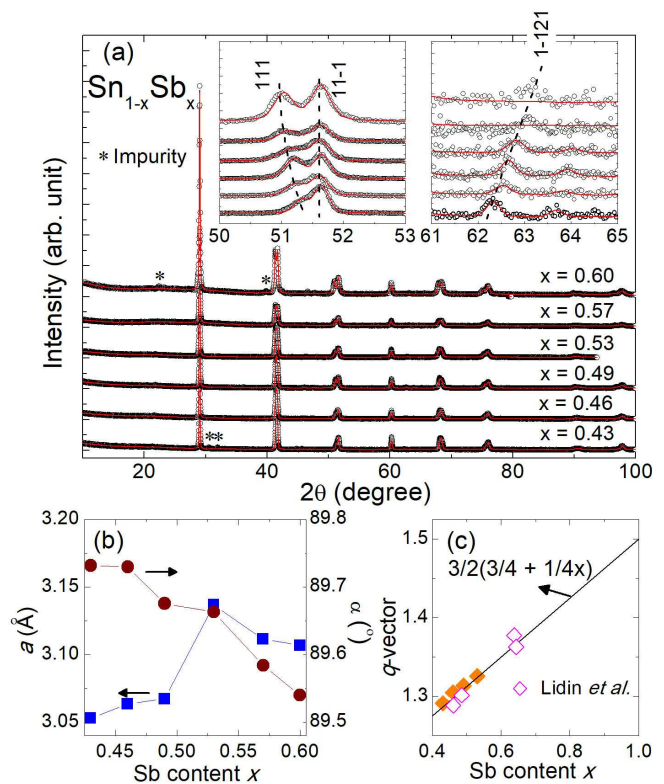}
\caption{(Color online)
(a) Powder x-ray diffraction patterns at room temperature (open symbols) and their refinement profiles (solid lines) for the IM-Sn$_{1-x}$Sb$_{x}$ samples with $x$ = 0.43, 0.46, 0.49, 0.53, 0.57, and 0.60. Note that the data are shifted vertically for clarity and the impurity phases for $x$ = 0.43 and 0.60 are marked by the asterisks. The two upper-right insets show zooms of the structural refinement results between 50$^{\circ}$ $\leq$ 2$\theta$ $\leq$ 53$^{\circ}$ and 61$^{\circ}$ $\leq$ 2$\theta$ $\leq$ 65$^{\circ}$, respectively. (b) and (c) show the refined lattice parameters and $q$-vector value plotted a function of the Sb content $x$, respectively. The open symbols and the solid line in (c) are the results from Ref.~\citenum{SnSbstructure2}.
}
\label{fig2}
\end{figure}

The XRD patterns together with structural refinement profiles for the series of IM-Sn$_{1-x}$Sb$_{x}$ samples are shown in Fig. 2(a), and the refined structure parameters are listed in Table I.
Here both Sn and Sb atoms occupy the same (0, 0, 0) position, and their occupancies are fixed by the stoichiometry.
For $x$ $\leq$ 0.53, there is a good agreement between the observed and calculated ones based on the $R$$\overline{3}$$m$(00$q$) superspace group, not only for the main peaks of the $R$$\overline{3}$m average structure but also for the satellite reflections due to the incommensurate modulation [see the inset of Fig. 2(a)]. At higher $x$ values, however, the satellite peaks becomes less distinct so that the $q$-value can not be determined. In addition to the main phase, a very small amount of Sn-rich and Sb-rich impurities are present for $x$ = 0.43 and 0.6, respectively, confirming that these compositions are located at the phase boundaries. Figure 2(b) and (c) show the $x$-dependence of the refined lattice parameters and value of $q$-vector, respectively. With increasing $x$, the $a$-axis shows a nonmonotonic behavior with a maximum at $x$ = 0.53. This is contrast to the case of Bi$_{1-x}$Sb$_{x}$, where the $a$-axis decreases monotonically with $x$ for 0 $\leq$ $x$ $\leq$ 1 \cite{Bi1-xSbxLP}, and suggests that Sb atoms do not simply substitute Sn atoms. Meanwhile, the $\alpha$ angle decreases slightly from $\sim$89.7$^{\circ}$ to $\sim$89.5$^{\circ}$, which is in accordance with the gradual increase in the splitting of (111)/(11-1) peaks and points to an enhanced rhombohedral distortion. On the other hand, the $q$-value shows a linear increase with $x$, and an extrapolation of these data to $x$ = 1 yields $q$ = 1.5, which is exactly the value for pure Sb. The overall results are well consistent with those reported previously \cite{SnSbstructure1,SnSbstructure2,SnSbSC}, and demonstrate the incommensurate ordering of Sn/Sb layers in IM-Sn$_{1-x}$Sb$_{x}$.

\begin{figure}
\includegraphics*[width=8.6cm]{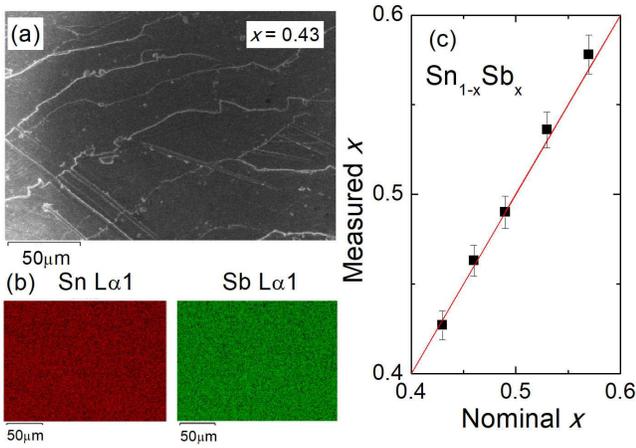}
\caption{(Color online)
(a) SEM image of the sample with $x$ = 0.43 taken at a magnification of 500X.
(b) EDX elemental mapping for the same sample.
(c) Comparison between the nominal and measured Sb contents $x$.
The solid line marks the equality between the two quantities.
}
\label{fig2}
\end{figure}

The chemical composition and homogeneity of these samples were checked by SEM and EDX measurements. As an example, the SEM image for the sample with $x$ = 0.43 is shown in Fig. 3(a).
The data reveal the presence of a series of steps and terraces, in line with the layered structure.
Remarkably, the EDX elemental mapping [Fig. 3(b)] results indicate that Sn and Sb distribute quite uniformly in the sample.
Note that this is also the case for other $x$ values (see Fig. S1 of the Supplemental Material\cite{SM}).
Hence, it appears that these samples are rather homogeneous in composition.
This clearly differs from a previous study of SnSb, where micro-phase segregation with a characteristic length of a few hundred $\mu$m was observed \cite{SnSbstructure1}.
Furthermore, as shown in Fig. 3(c), the nominal and measured Sb contents show good agreement within the experimental error, further attesting the high sample quality.

\subsection{Resistivity and magnetic susceptibility}
Figure 3 shows the temperature dependence of resistivity for the IM-Sn$_{1-x}$Sb$_{x}$ samples. For better comparison, the data for each $x$ are normalized to the value at 300 K.
A metallic behavior is observed in all cases, while the residual resistivity ratio (RRR) decreases with increasing $x$.
Note that the Hall coefficient at 1.8 K remains negative and nearly constant at a value of $-$2.5 $\times$ 10$^{-4}$ cm$^{3}$/C across this $x$ range (see Fig. S2 of the Supplemental Material\cite{SM}).
It thus seems that the incorporation of more Sb atoms does not introduce additional charge carriers but results in stronger electron scattering.
At low temperature, all the samples show a transition to to a zero resistivity state, indicating the occurrence of SC.
Following Ref.~\citenum{SnSbSC}, we define the resistive transition temperature $T_{\rm c}^{\rho}$ as the temperature corresponding to the midpoint of the resistive transition, and find that it increases monotonically with $x$ from 2.06 to 3.29 K. Nevertheless, since the previous study on SnSb shows that $T_{\rm c}^{\rho}$ is not a bulk property\cite{SnSbSC}, its evolution with $x$ is of extrinsic origin, to which we will return below.
\begin{figure}
\includegraphics*[width=8.2cm]{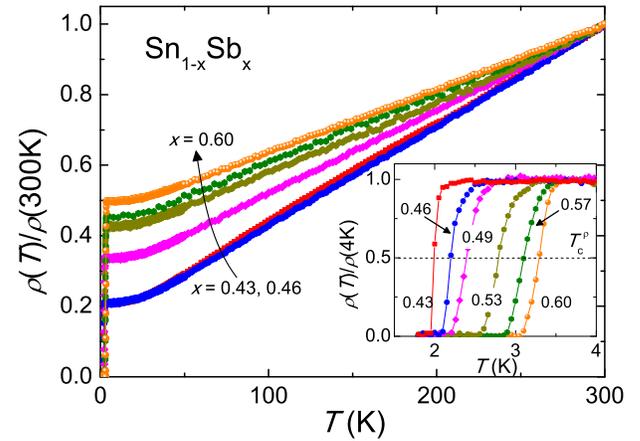}
\caption{(Color online)
Temperature dependence of resistivity normalized to its value at 300 K for the IM-Sn$_{1-x}$Sb$_{x}$ samples.
The lower-right inset shows a zoom of the low-temperature data normalized to the value at 4 K.
The dashed line denotes the midpoint of the resistive transition.
}
\label{fig3}
\end{figure}

\begin{figure}
\includegraphics*[width=8.2cm]{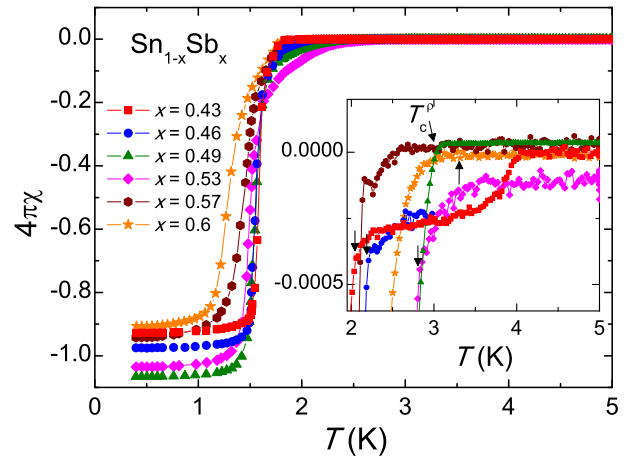}
\caption{(Color online)
Temperature dependence of magnetic susceptibility measured under an applied field of 5 Oe for the IM-Sn$_{1-x}$Sb$_{x}$ samples. The inset shows a zoom of the data near the onset of diamagnetic transition.
The resistive transition temperatures $T_{\rm c}^{\rho}$ are marked by the arrows. Note that $T_{\rm c}^{\rho}$ of 2.5 K for $x$ = 0.49 is out of the range of the figure and hence not indicated.
}
\label{fig4}
\end{figure}
Figure 4 shows the zero-field cooling (ZFC) magnetic susceptibility below 5 K for the IM-Sn$_{1-x}$Sb$_{x}$ samples measured under an applied field of 5 Oe.
The demagnetization effect was corrected by assuming that the grains have a cubic shape with a demagnetization factor of 0.3.
Indeed, a diamagnetic response is observed for all cases, and the shielding fractions (SF) at 0.4 K are in the range between 91\% and 106\%, clearly indicating that all the samples are bulk superconductors.
However, as in the case of SnSb \cite{SnSbSC}, the initial transition is rather broad, and the SF at $T_{\rm c}^{\rho}$ is less than 0.5\% for these samples except for $x$ = 0.49. In particular, as shown in the inset of Fig. 4, no diamagnetic transition is visible for $x$ = 0.57 and 0.6.
These results strongly suggest that the resistive transition involves only a very small fraction of the sample volume. In passing, there exists another
diamagnetic transition around 4 K for $x$ = 0.43, which is presumably ascribed to the Sn-based solid solution observed by the XRD.

\subsection{Specific heat}
\begin{figure}
\includegraphics*[width=7cm]{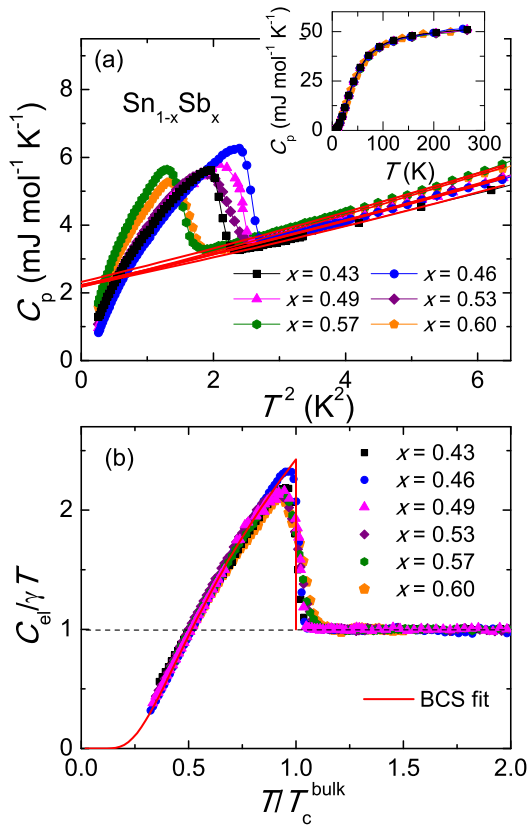}
\caption{(Color online)
(a) Low-temperature specific heat plotted as $C_{\rm p}$/$T$ versus $T^{2}$ for the IM-Sn$_{1-x}$Sb$_{x}$ samples.
The solid lines are fits by the Debye model to the data.
The upper right inset shows the $C_{\rm p}$ data up to 270 K.
(b) Normalized electronic specific heat data for these samples.
The solid line denotes the curve given by the weak-coupling BCS theory.
}
\label{fig5}
\end{figure}
To further characterize the IM-Sn$_{1-x}$Sb$_{x}$ samples,
we have performed specific heat ($C_{\rm p}$) measurements in the temperature range between 0.5 and 300 K.
The low temperature data are plotted as $C_{\rm p}$/$T$ versus $T^{2}$ in Fig. 5(a).
For all samples, a clear $C_{\rm p}$ jump is observed, corroborating the bulk nature of SC.
On the other hand, the $C_{\rm p}$ data in the normal state can be well described by the Debye model $C_{\rm p}$/$T$ = $\gamma$ + $\beta_{3}$$T^{2}$ + $\beta_{5}$$T^{4}$, where $\gamma$ and $\beta_{3}$/$\beta_{5}$ are the electronic and phonon specific heat coefficients, respectively. With $\beta_{3}$ known, the Debye temperature $\Theta_{\rm D}$ is then calculated as $\Theta_{\rm D}$ = (12$\pi^{4}$$N$$R$/5$\beta_{3}$)$^{1/3}$, where $N$ = 2 is the number of atoms per unit cell and $R$ = 8.314 J mol$^{-1}$ K$^{-1}$ is the molar gas constant.
As can be seen in Table I, with increasing $x$, the $\gamma$ value remains nearly unchanged while $\Theta_{\rm D}$ decreases monotonically.
Since $\Theta_{\rm D}$ is inversely proportional to $\sqrt{M}$ ($M$ is the molar mass), its decrease with $x$ is as expected.
Nevertheless, the change in $\Theta_{\rm D}$ is small such that the $C_{\rm p}$ data of these samples almost overlap with each other at high temperature [see the inset of Fig. 5(a)].

Figure 5(b) shows the normalized electronic specific heat $C_{\rm el}$/$\gamma$$T$ plotted as a function of $T$/$T_{\rm c}^{\rm bulk}$, where $C_{\rm el}$/$T$ = $C_{\rm p}$/$T$ $-$ $\beta_{3}$$T^{2}$ $-$ $\beta_{5}$$T^{4}$ and the bulk superconducting transition temperature $T_{\rm c}^{\rm bulk}$ is determined from the entropy balance.
The $T_{\rm c}^{\rm bulk}$ values are 1.45 K, 1.58 K, 1.53 K, 1.47 K, 1.22 K, 1.23 K for $x$ = 0.43, 0.46, 0.49, 0.53, 0.57, 0.60, respectively, and indeed significantly lower than the corresponding $T_{\rm c}^{\rho}$.
Remarkably, despite the difference in $T_{\rm c}^{\rm bulk}$, all the normalized data collapse on the same curve, suggesting that the superconducting gap structure remains unchanged with varying $x$.
Furthermore, as found previously \cite{SnSbSC}, the $C_{\rm el}$/$\gamma$$T$ jump follows closely that of the weak-coupling Bardeen-Copper-Schrieffer (BCS) theory \cite{BCS}, and hence points to a fully gapped superconducting state in these samples. Assuming that the superconducting pairing is phonon mediated, one can calculate the electron-phonon coupling constant $\lambda_{\rm ep}$ using the inverted McMillan formula \cite{Mcmillan},
\begin{equation}
\lambda_{\rm ph} = \frac{1.04 + \mu^{\ast} \rm ln(\Theta_{\rm D}/1.45\emph{T}_{\rm c})}{(1 - 0.62\mu^{\ast})\rm ln(\Theta_{\rm D}/1.45\emph{T}_{\rm c}) - 1.04},
\end{equation}
where $\mu^{\ast}$ is the Coulomb repulsion pseudopotential. Taking $\mu^{\ast}$ = 0.13, $\lambda_{\rm ep}$ values of 0.5-0.52 are obtained.
This suggests that these IM-Sn$_{1-x}$Sb$_{x}$ samples are weakly coupled superconductors, in line with the magnitude of $C_{\rm el}$/$\gamma$$T$ jumps.
\begin{table}
\caption{Structural and physical parameters of the IM-Sn$_{1-x}$Sb$_{x}$ samples.}
\renewcommand\arraystretch{1.3}
\begin{tabular}{p{2cm}<{\centering}p{0.9cm}<{\centering}p{0.9cm}<{\centering}p{0.9cm}<{\centering}p{0.9cm}<{\centering}p{0.9cm}<{\centering}p{0.9cm}<{\centering}}
\\
\hline 
   &  0.43   &   0.46  & 0.49 & 0.53 &  0.57 &  0.60\\

\hline 
$a$ ({\AA})							& 	 3.053 	 & 3.063	& 3.067 & 3.137 & 3.111 & 3.107	\\
$\alpha$ ($^{\circ}$)							& 	  89.73 	 & 89.73 & 89.68 & 89.66 & 89.58 & 89.54		\\
$q$-vector 				&      1.291    & 			1.305 & 1.313 & 1.325 & $-$ & 	$-$		 \\
$R_{\rm wp}$	&      7.8\% &   7.2\%  & 9.2\% & 9.6\% & 9.3\% & 11.3\% \\
$R_{\rm p}$ 	&       5.4\% & 5.3\% & 6.3\% & 6.9\% & 6.6\% & 7.5\% \\
$T_{\rm c}^{\rho}$ (K)							& 	  2.06 	 & 2.22	& 2.40 & 2.77 & 3.11 & 3.29	\\
$T_{\rm c}^{\rm bulk}$ (K)							& 	  1.45 	 & 1.58 & 1.53 & 1.47 & 1.22 & 1.23		\\
$\Theta_{\rm D}$ (K)				&      212    & 			207 & 205 & 203 & 201 & 	199		 \\
$\gamma$ (mJ mol$^{-1}$ K$^{-2}$)	&       2.19 & 2.22 & 2.22 & 2.18 & 2.32 & 2.19 \\
$\lambda_{\rm ph}$	&      0.51 &   0.52  & 0.52 & 0.52 & 0.50 & 0.50 \\

\hline
\hline 
\end{tabular}
\label{Table3}
\end{table}

\subsection{Superconducting phase diagram}
\begin{figure}
\includegraphics*[width=8.6cm]{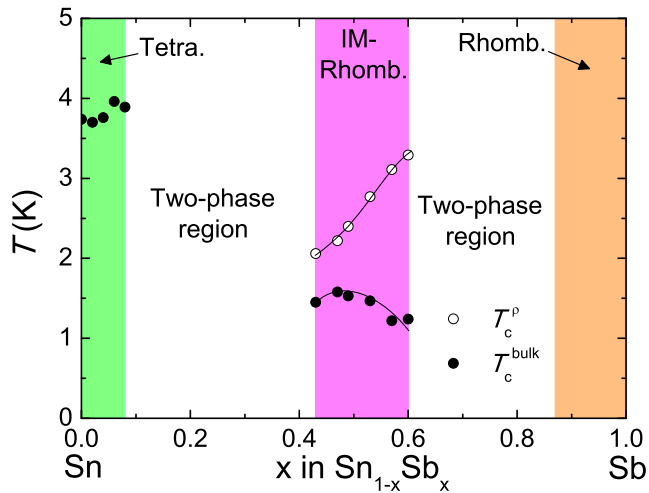}
\caption{(Color online)
$T_{\rm c}$ versus $x$ phase diagram for the binary Sn-Sb system.
The $T_{\rm c}^{\rm bulk}$ data for $x$ $\leq$ 0.08 are taken from Ref.~\citenum{diluteSn1-xSbx}.
}
\label{fig4}
\end{figure}
Figure 6 shows the $T_{\rm c}$ versus composition phase diagram for the binary Sn-Sb system, which incorporates the present results with those from Ref.~\citenum{diluteSn1-xSbx}.
As can be seen, our results reveal the presence of a second superconducting region for the system, which is structurally different from the known one based on tetragonal Sn.
Although $T_{\rm c}^{\rm bulk}$ of IM-Sn$_{1-x}$Sb$_{x}$ is considerably lower than that of the Sn-based solid solution,
it shows a dome-like dependence on $x$ with a maximum around $x^{\rm opt}$ = 0.46, as can be directly seen in Fig. 5(a).
Since the carrier concentration and electronic specific heat coefficient are nearly $x$-independent, this behavior is unlikely
related to a change in the density of states at the Fermi level. It is prudent to note that the most Sn-rich composition $x$ = 0.43 is built up solely by the [Sn$_{4}$Sb$_{3}$] septuple layers, and the stoichiometry for $x$ $>$ 0.43 is kept through the intercalation of additional Sb layers \cite{SnSbPD2}. Hence it appears that the [Sn$_{4}$Sb$_{3}$] layer is
the essential building block for SC in IM-Sn$_{1-x}$Sb$_{x}$. On one hand, the Sb-layer intercalation increases the separation between the [Sn$_{4}$Sb$_{3}$] blocks, which may enhance $T_{\rm c}$. On the other hand, as shown above, the increase of Sb content results in a decrease in $\Theta_{\rm D}$ and a stronger electron scattering, both of which tend to suppress $T_{\rm c}^{\rm bulk}$. The competition between these two effects is the likely cause for the nonmonotonic $x$-dependence of $T_{\rm c}^{\rm bulk}$.

Another prominent feature seen from Fig. 6 is the monotonic increase of $T_{\rm c}^{\rho}$.
Notably, the difference between $T_{\rm c}^{\rho}$ and $T_{\rm c}^{\rm bulk}$ gets larger with increasing $x$, and their ratio amounts to $\sim$2.7 at $x$ = 0.6.
As shown above, the zero resistivity transition in IM-Sn$_{1-x}$Sb$_{x}$ is confined to a very small part of the sample, and the possible origins include compositional inhomogeneity and strain effect at the grain boundaries \cite{SnSbSC}.
The contrast behavior between $T_{\rm c}^{\rho}$ and $T_{\rm c}^{\rm bulk}$, in both magnitude and variation with $x$, rules out the former possibility.
This is also corroborated by the SEM-EDX results shown above.
As for the latter one, it is worth mentioning that an enhanced $T_{\rm c}$ possibly due to internal strain was observed in ($\gamma$-Sn + $\beta$-InSn) composite alloy \cite{compositealloy}.
However, this $T_{\rm c}$ value is found to be composition independent \cite{compositealloy}, which also differs from the present case.
In this regard, further investigations of the grain boundary properties in IM-Sn$_{1-x}$Sb$_{x}$
would be necessary to better understand the observed $T_{\rm c}^{\rho}$ behavior.

\subsection{Band structure calculations}
\begin{figure}
\includegraphics*[width=8.6cm]{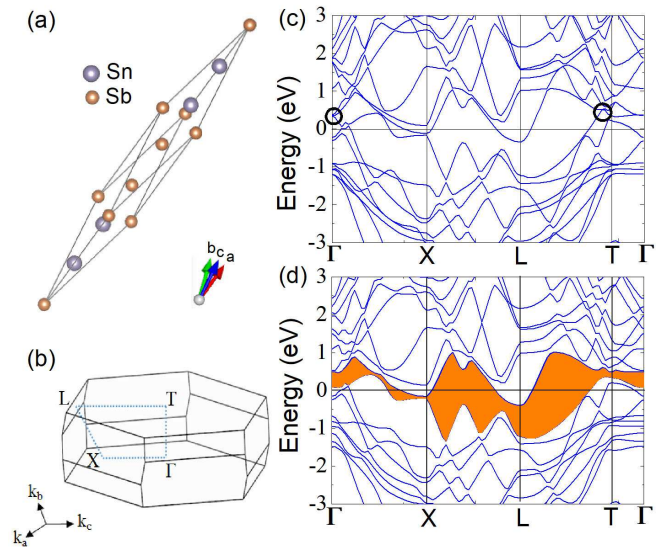}
\caption{(Color online)
(a) and (b) show the primitive unit cell and first Brillouin zone of rhombohedral Sn$_{4}$Sb$_{3}$.
The four TRIM points are labeled.
The calculated band structure plots along $\Gamma$-X-L-T-$\Gamma$ direction of the Brilloin zone without and with SOC are shown in (c) and (d).
The circles in (c) are a guide to the eyes.
The orange area in (d) denotes the continuous gap separating the valence and conduction bands.
}
\label{fig5}
\end{figure}
From the above results, one can see that the properties of IM-Sn$_{1-x}$Sb$_{x}$ are a combination of those of its constituent elements: it adopts the same structure as Sb, and shows SC in analogy to Sn.
Given that Sb is topologically nontrivial \cite{paritycounting}, it is of natural interest to examine the band topology of IM-Sn$_{1-x}$Sb$_{x}$.
To provide insight into this issue, we calculated the band structure of the most representative composition Sn$_{4}$Sb$_{3}$ ($x$ = 0.43).
The crystal structure model\cite{SnSbstructure2} used in the calculation and the corresponding Brillouin zone are shown in Fig. 7
(a) and (b), respectively.
For Sn$_{4}$Sb$_{3}$, there are four time-reversal-invariant momenta (TRIM) points, namely $\Gamma$, X, L, T.
As can be seen in Fig. 7(c), the calculation results without SOC indicate that Sn$_{4}$Sb$_{3}$ is a metallic system with both conduction and valence bands crossing the Fermi level.
However, when turning on the SOC [Fig. 7(d)], the valence and conduction bands become separated over the whole Brillouin zone so that one can proceed to calculate the Z$_{2}$ topological invariants.
With the parity counting method \cite{paritycounting}, the calculated parity eigenvalues are $-$, $-$, +, +  for the $\Gamma$, X, L, T points, respectively, which yields the four Z$_{2}$ invariants ($\nu_{0}$;$\nu_{1}$,$\nu_{2}$,$\nu_{3}$) = (0;111).
This is further verified by calculating the Wilson loop, which tracks the evolution lines of Wannier charge centers on six TRIM planes (see Fig. S3 of the Supplemental Material\cite{SM}).
These results indicate that Sn$_{4}$Sb$_{3}$ partially inherits the nontrivial band topology of Sb [(1;111)], and is in the weak TI phase. In this case, there is an even number of surface states for specific surface terminations \cite{weakTI}.
Although these surface states are susceptible to localization by disorder, they can evolve into topologically protected one dimensional (1D) helical modes along the dislocation lines\cite{1Dhelicalmodes}.

There are two immediate implications of this result. Firstly, these 1D helical modes may support topological SC due to the proximity effect from the bulk.
As a consequence, Majorana fermions confined to the dislocation lines could be detected by tunneling experiments \cite{STM1,STM2}.
Secondly, it is reasonable to speculate that IM-Sn$_{1-x}$Sb$_{x}$ for the whole $x$ range is topologically nontrivial, since the insertion of Sb layers between the [Sn$_{4}$Sb$_{3}$] blocks does not induce a significant change in the band structure near the Fermi level, as evidenced by the transport and thermodynamic measurements. Nevertheless, given that Sn$_{4}$Sb$_{3}$ and Sb belong to different topological classes, the Z$_{2}$ invariants may vary with increasing $x$. In the respect, the elucidation of the interplay between the structural modulation and band topology in IM-Sn$_{1-x}$Sb$_{x}$ could be a focus of future research.

\section{Conclusion}
In summary, we have presented a combined experimental and theoretical study of Sn$_{1-x}$Sb$_{x}$ alloy with 0.43 $\leq$ $x$ $\leq$ 0.6, which exhibits an incommensurate structural modulation due to the formation of a natural superlattice of [Sn$_{4}$Sb$_{3}$] and Sb layers.
BCS-like SC is observed for the whole $x$ range, although a zero resistivity transition takes place well above the bulk superconducting transition.
The established phase diagram shows that the bulk $T_{\rm c}$ has a maximum of 1.58 K at $x$ = 0.46, though the Hall and electronic specific coefficients remain nearly constant with varying $x$.
This nonmonotonic dependence of bulk $T_{\rm c}$ is tentatively attributed to the competing effects induced by the insertion of Sb layers between the [Sn$_{4}$Sb$_{3}$] lamellae, which are the essential building blocks for SC.
On the other hand, the resistive $T_{\rm c}$ increases monotonically with $x$ from 2.06 K to 3.29 K.
This, along with an uniform distribution of Sn and Sb in these alloys found by EDX elemental mapping, suggests that the resistive transition results from the strain effect that develops at the grain boundaries rather than compositional inhomogeneity.
First-principles calculations on the band structure of the most Sn-rich composition and representative composition Sn$_{4}$Sb$_{3}$ shows that it is topologically nontrivial with Z$_{2}$ invariants of (0;111).
Consequently, Majorana fermions may exist along its dislocation lines due to the superconducting proximity effect from the bulk.
Our results not only identify a new superconducting region in the Sn-Sb phase diagram, but also supply the first alloy system that combines structural modulation, nontrivial band topology and SC, which opens the door to study the interplay between these properties.
\\

\section*{ACKNOWLEGEMENT}
This work is financially supported by the National Key Research Development Program of China (No.2017YFA0303002), the National Natural Science Foundation of China (No.61574123), the Fundamental Research Funds for the Central Universities of China (2019FZA3004) and Zhejiang Provincial Natural Science Foundation (D19A040001).


\begin{thebibliography}{99}
\expandafter\ifx\csname natexlab\endcsname\relax\def\natexlab#1{#1}\fi
\expandafter\ifx\csname bibnamefont\endcsname\relax
  \def\bibnamefont#1{#1}\fi
\expandafter\ifx\csname bibfnamefont\endcsname\relax
  \def\bibfnamefont#1{#1}\fi
\expandafter\ifx\csname citenamefont\endcsname\relax
  \def\citenamefont#1{#1}\fi
\expandafter\ifx\csname url\endcsname\relax
  \def\url#1{\texttt{#1}}\fi
\expandafter\ifx\csname urlprefix\endcsname\relax\def\urlprefix{URL }\fi
\providecommand{\bibinfo}[2]{#2}
\providecommand{\eprint}[2][]{\url{#2}}

\bibitem{kaneRMP}
M. Z. Hasan and C. L. Kane,
Rev. Mod. Phys. {\bf 82}, 3045(2010).

\bibitem{QiRMP}
X. L. Qi and S. C. Zhang,
Rev. Mod. Phys. {\bf 83}, 1057 (2011).

\bibitem{paritycounting}
L. Fu and C. L. Kane,
Phys. Rev. B {\bf 76}, 045302 (2007).

\bibitem{Bi1-xSbxarpes}
D. Hsieh, D. Qian, L. Wray, Y. Xia, Y. S. Hor, R. J. Cava, and M. Z. Hasan,
Nature {\bf 452}, 970 (2008).

\bibitem{Bi1-xSbxalloy}
J. C. Y. Teo, L. Fu, and C. L. Kane,
Phys. Rev. B {\bf 78}, 045426 (2008).

\bibitem{AliRPP}
J. Alicea,
Rep. Prog. Phys. {\bf 75}, 076501 (2012).

\bibitem{AndoRPP}
M. Sato and Y. Ando,
Rep. Prog. Phys. {\bf 80}, 076501 (2017).

\bibitem{CuxBi2Se3Hor}
Y. S. Hor, A. J.Williams, J. G. Checkelsky, P.  Roushan, J. Seo, Q. Xu, H. W. Zandbergen, A. Yazdani, N. P. Ong, and R. J. Cava,
Phys. Rev. Lett. {\bf 104}, 057001 (2010).

\bibitem{CuxBi2Se3sasaki}
S. Sasaki, M. Kriener, K. Segawa, K. Yada, Y. Tanaka, M. Sato, and Y. Ando,
Phys. Rev. Lett. {\bf 107}, 217001 (2011).

\bibitem{SrxBi2Se3}
Z. H. Liu, X. Yao, J. F. Shao, M. Zuo, L. Pi, S. Tan, J. S. Zhang, and Y. H. Zhang,
J. Am. Chem. Soc. {\bf 137}, 10512 (2015).

\bibitem{NbxBi2Se3}
T. Asaba, B. J. Lawson, C. Tinsman, L. Chen, P. Corbae, G. Li, Y. Qiu, Y. S. Hor, L. Fu, and L. Li,
Phys. Rev. X {\bf 7}, 011009 (2017).

\bibitem{Sn1-xInxTe}
S. Sasaki, Z. Ren, A. A. Taskin, K. Segawa, L. Fu, and Y. Ando,
Phys. Rev. Lett. {\bf 109}, 217004 (2012).

\bibitem{Tl5Te3}
K. E. Arpino, D. C. Wallace, Y. F. Nie, T. Birol, P. D. C. King, S. Chatterjee, M. Uchida, S. M. Koohpayeh, J. J. Wen, K. Page, C. J. Fennie, K. M. Shen, and T. M. McQueen,
Phys. Rev. Lett. {\bf 112}, 017002 (2014).

\bibitem{PbTaSe2}
M. N. Ali, Q. D. Gibson, T. Klimczuk, and R. J. Cava,
Phys. Rev. B {\bf 89}, 020505(R) (2014).

\bibitem{Au2Pb}
L. M. Schoop, L. S. Xie, R. Chen, Q. D. Gibson, S. H. Lapidus, I. Kimchi, M. Hirschberger, N. Haldolaarachchige, M. N. Ali, C. A. Belvin, T. Liang, J. B. Neaton, N. P. Ong, A. Vishwanath, and R. J. Cava,
Phys. Rev. B {\bf 91}, 214517 (2015).

\bibitem{Tl0.6Bi2Te3}
Z. W. Wang, A. A. Taskin, T. Frolich, M. Braden, and Y. Ando,
Chem. Mater. {\bf 28}, 779 (2016).

\bibitem{PdBi2}
Y. F. Lv, W. L. Wang, Y. M. Zhang, H. Ding, W. Li, L. L. Wang, K. He, C. L. Song, X. C. Ma, and Q. K. Xue,
Science Bulletin {\bf 62}, 852 (2017).

\bibitem{SnSbPD1}
C. Schmetterer, J. Polt, and H. Flandorfer,
J. Alloys Compd. {\bf 728}, 497 (2018).

\bibitem{SnSbPD2}
C. Schmetterer, J. Polt, and H. Flandorfer,
J. Alloys Compd. {\bf 523}, 523 (2018).

\bibitem{SnSbstructure1}
L. Noren, R. L. Withers, S. Schmid, F. J. Brink, and V. Ting,
J. Solid State Chem. {\bf 179}, 404 (2006).

\bibitem{SnSbstructure2}
S. Lidin, J. Christensen, K. Jasson, D. Fredrickson, R. Withers, L. Noren, and S. Schmid,
Inorg. Chem. {\bf 48}, 5497 (2009).

\bibitem{SnSbEP1}
J. Yang, M. Winter, and J. O. Besenhard,
Solid State Ion. {\bf 90}, 281 (1996).

\bibitem{SnSbEP2}
J. Yang, M. Wachtler, M. Winter, and J. O. Besenhard,
Solid State Lett. {\bf 2}, 161 (1999).

\bibitem{SnSbEP3}
H. Li, Z. X. Wang, L. Q. Chen, and J. X. Huang,
Adv. Mater. {\bf 21}, 4593 (2009).

\bibitem{SnSbEP4}
J. H. Choi, C. W. Ha, H. Y. Choi, J. W. Seong, C. M. Park, and S. M. Lee,
J. Power Sources {\bf 386}, 34 (2018).

\bibitem{SnSbSC}
B. Liu, J. F. Wu, Y. W. Cui, H. D. Wang, Y. Liu, Z. C. Wang, Z. Ren, and G. H. Cao,
Supercond. Sci. Technol. {\bf 31}, 125011 (2018).

\bibitem{JANA}
V. Petricek, M. Dusek, and L. Palatinus,
{\it Z. Kristallogr} {\bf 229}, 345 (2014).

\bibitem{VASP}
G. Kresse and J. Hafner,
Phys. Rev. B {\bf 47}, 558 (1993).

\bibitem{pseudopotential}
P. E. Blochl,
Phys. Rev. B {\bf 50}, 17953 (1994).

\bibitem{mesh}
H. J. Monkhorst and J. D. Pack,
Phys. Rev. B {\bf 13}, 5188 (1976).

\bibitem{wilsonloop}
R. Yu, X. L. Qi, A. Bernevig, Z. Fang, and X. Dai,
Phys. Rev. B {\bf 84}, 0075119 (2011).

\bibitem{Bi1-xSbxLP}
H. Berger, B. Christ, and J. Troschke,
Crystal Res. Technol. {\bf 17}, 1233 (1982).

\bibitem{SM}
See Supplemental Material at (link to be inserted) for more SEM-EDX data, Hall and theoretical calculation results.

\bibitem{BCS}
J. Bardeen, L. N. Copper, and J. R. Schreiffer,
{\it Phys. Rev.} {\bf 108}, 1175 (1957).

\bibitem{Mcmillan}
W. L. McMillan,
Phys. Rev. {\bf 167}, 331 (1968).

\bibitem{diluteSn1-xSbx}
W. F. Love,
Phys. Rev. {\bf 92}, 238 (1953).

\bibitem{compositealloy}
A. C. Gandi and S. Y. Wu,
Inorg. Chem. {\bf 58}, 794 (2019).

\bibitem{weakTI}
L. Fu, C. L. Kane, and E. J. Mele,
Phys. Rev. Lett. {\bf 98}, 106803 (2007).

\bibitem{1Dhelicalmodes}
K. I. Imura, Y. Takane, and A. Tanaka,
Phys. Rev. B {\bf 84}, 035443 (2011).

\bibitem{STM1}
S. Nadj-Perge, I. K. Drozdov, J. Li, H. Chen, S. J. Jeon, A. H. MacDonald, B. A. Bernevig, and A. Yazdani,
Science {\bf 346}, 602 (2014).

\bibitem{STM2}
D. Chevallier and J. Klinovaja,
Phys. Rev. B {\bf 94}, 035417 (2016).

\end{thebibliography}
\end{document}